\documentclass[aps,prb,twocolumn,10pt]{revtex4-2}
\usepackage{libertine}
\usepackage[T1]{fontenc}
\usepackage[libertine,upint]{newtxmath}
\usepackage[cal=stixtwofancy,frak=stixtwo]{mathalfa}
\usepackage{graphicx,microtype,mathtools,bm,caption,subcaption,floatrow,xcolor}
\definecolor{linkscolor}{cmyk}{0.6, 0.3, 0, 0.9}
\usepackage[colorlinks=true,allcolors=linkscolor!60]{hyperref}

\DeclarePairedDelimiterX\braket[2]{\langle}{\rangle}{#1 \delimsize\vert #2}
\DeclarePairedDelimiterX\matrixel[3]{\langle}{\rangle}{#1 \delimsize\vert #2 \delimsize\vert #3}
\DeclarePairedDelimiter\abs{\lvert}{\rvert}

\DeclareMathOperator{\sign}{sign}
\DeclareMathOperator{\const}{const}

\DeclareMathOperator{\im}{Im}

\NewDocumentCommand{\grad}{e{_^}}{%
    \mathop{}\!
    \nabla
    \IfValueT{#1}{_{\!#1}}
    \IfValueT{#2}{^{#2}}
}

\begin{document}

\title{Fermi Surface Geometry and Optical Conductivity of a 2D Electron Gas\\ near an Ising-Nematic Quantum Critical
Point}

\author{Yasha Gindikin}
\author{Andrey V.\ Chubukov}
\affiliation{W.I.\ Fine Theoretical Physics Institute and School of Physics and Astronomy, University of Minnesota,
Minneapolis, MN 55455, USA}

\begin{abstract}
We analyze optical conductivity of a clean two-dimensional electron system in a Fermi liquid regime near a $T=0$ Ising-nematic quantum critical point (QCP), and extrapolate the results to a QCP\@. We employ direct perturbation theory up to the two-loop order to elucidate how the Fermi surface's geometry (convex vs.\ concave) and fermionic dispersion (parabolic vs.\ non-parabolic) affect the scaling of the optical conductivity, $\sigma(\omega)$, with frequency $\omega$ and correlation length $\xi$. We find that for a convex Fermi surface the leading terms in the optical conductivity cancel out, leaving a sub-leading contribution  $\sigma (\omega) \propto \omega^2 \xi^4 \mathcal{L}$, where $\mathcal{L} =\const$ for a parabolic dispersion and $\mathcal{L} \propto \log{\omega \xi^3}$ in a generic case. For a concave Fermi surface, the leading terms do not cancel, and $\sigma (\omega) \propto \xi^2$. We extrapolate these results to a QCP and obtain $\sigma (\omega) \propto \omega^{2/3}$ for a convex Fermi surface and $\sigma (\omega) \propto 1/\omega^{2/3}$ for a concave Fermi surface.
\end{abstract}

\maketitle
\section{Introduction}
The study of quantum dynamics of electrons near an electron-driven quantum phase transition at $T=0$ offers a fascinating window into the intricate interplay between electronic degrees of freedom and bosonic fluctuations of the order parameter. In this work, we consider the optical conductivity of electrons near a $T=0$ nematic transition, which leads to quadrupolar distortion of the Fermi surface (FS)~\cite{doi:10.1146/annurev-conmatphys-070909-103925}. There are two major scenarios of nematicity. The first postulates that nematicity emerges in proximity to an ordered state that breaks lattice rotational symmetry along with another symmetry, e.g., spin-rotational one~\cite{Chandra90,*Sachdev08,*Xu_Qi09}. In this scenario, a nematic state is a partially ordered state wherein spin-rotational symmetry breaks prior to other symmetries. This scenario has been successfully applied to a number of $\mathrm{FeAs}$ systems, where nematic order appears in the vicinity of a stripe magnetic order~\cite{RMF12,*Fernandes_2017}. Another scenario postulates that nematic order is the result of  a Pomeranchuk instability in the channel with angular momentum $l=2$~\cite{pomeranchuk1958stability}. This  scenario has been applied to $\mathrm{FeSe}$, where a nematic order is not accompanied by a stripe magnetic order~\cite{Bascones09,*Phillips10,*PhysRevX.6.041045}. Our work explores the second scenario. We discuss optical conductivity — a highly powerful experimental tool to extract information about the behavior of correlated electrons~\cite{armitage2018electrodynamics,RevModPhys.83.471} — near the onset of $l=2$ charge Pomeranchuk order in a two-dimensional (2D) metal.

At a nematic quantum-critical point (QCP), collective bosonic excitations are massless and Landau overdamped. In 2D, the interaction between these excitations and low-energy fermions destroys a Fermi liquid, resulting in a fermionic self-energy $\Sigma (\omega) \propto \omega^{2/3}$ (see, e.g., Refs.~\cite{metzner0,*DellAnna2006,*Oganesyan2001,*Garst2010,*avi}). Similar behavior occurs in fermions coupled to a $U(1)$ gauge field~\cite{PALee1989,*Monien1993,*Nayak1994,*Altshuler1994}, and near a ferromagnetic QCP~\cite{PhysRevB.74.195126,*pepin1}, provided that ferromagnetism is of the Ising type; otherwise, a continuous $q=0$ transition would be precluded by non-analytic terms in the free energy~\cite{maslov2}. The behavior of conductivity is rather tricky. For interaction with a small momentum transfer there is no Umklapp scattering~\cite{PhysRevLett.106.106403}, unless the FS touches Brillouin zone boundary~\cite{palee2021}, hence for a clean system with small enough FS static conductivity is infinite. Nonetheless, optical conductivity is generally finite~\cite{Maslov_2017}. The conductivity (more precisely, its real part, $\sigma (\omega)$), is related to the retarded current-current correlator $\Pi (\omega)$ via the Kubo formula
\begin{equation}
	\label{eq:a}
	\sigma (\omega) \propto \im \frac{\Pi (\omega + i \delta)}{\omega}\,.
\end{equation}

\begin{figure}[htb]
	\centering
	\begin{subfigure}{.5\linewidth}
	  \centering
	  \includegraphics[width=.85\linewidth]{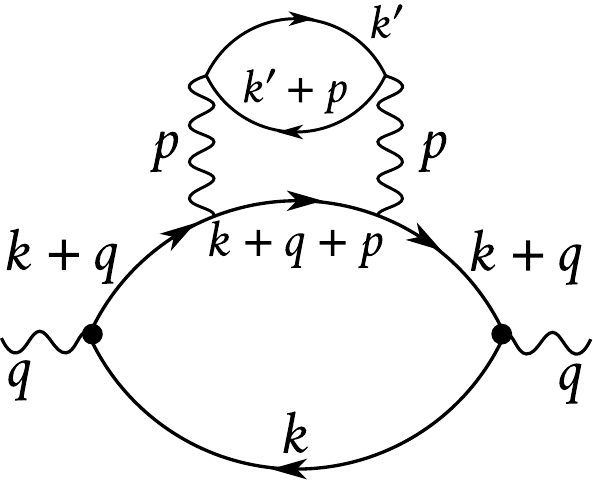}
	  \caption{
	  \label{fig:se1}}
	\end{subfigure}%
	\hfill
	\begin{subfigure}{.5\linewidth}
	  \centering
	  \includegraphics[width=.85\linewidth]{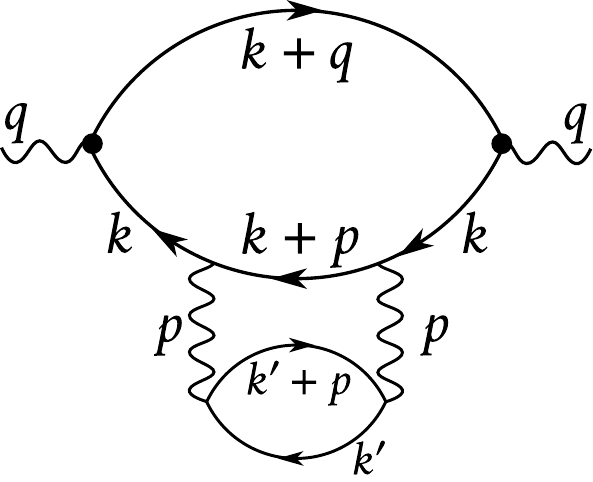}
	 \caption{
	  \label{fig:se2}}
	\end{subfigure}

	\bigskip
	\begin{subfigure}{.5\linewidth}
	  \centering
	  \includegraphics[width=0.85\linewidth]{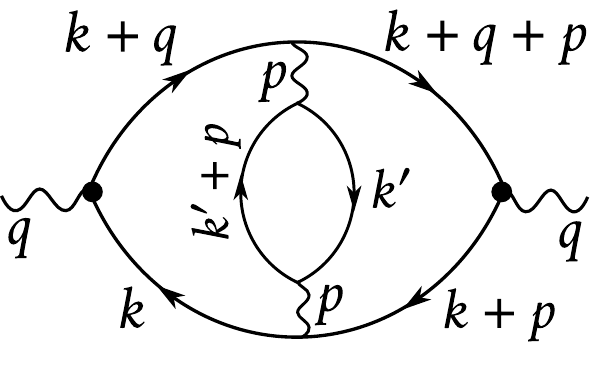}
	  \caption{
	  \label{fig:MT}}
	\end{subfigure}

	\bigskip
	\begin{subfigure}{.5\linewidth}
	  \centering
	  \includegraphics[width=0.95\linewidth]{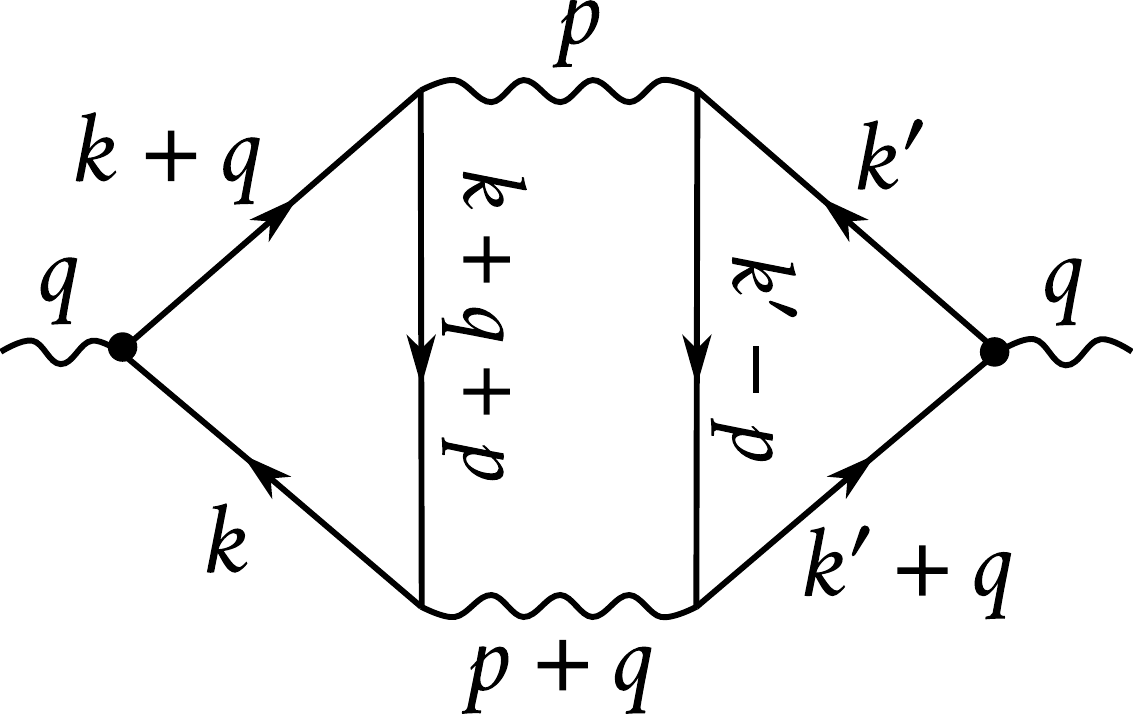}
	  \caption{
	  \label{fig:AL1}}
	\end{subfigure}%
	\begin{subfigure}{.5\linewidth}
	  \centering
	  \includegraphics[width=0.95\linewidth]{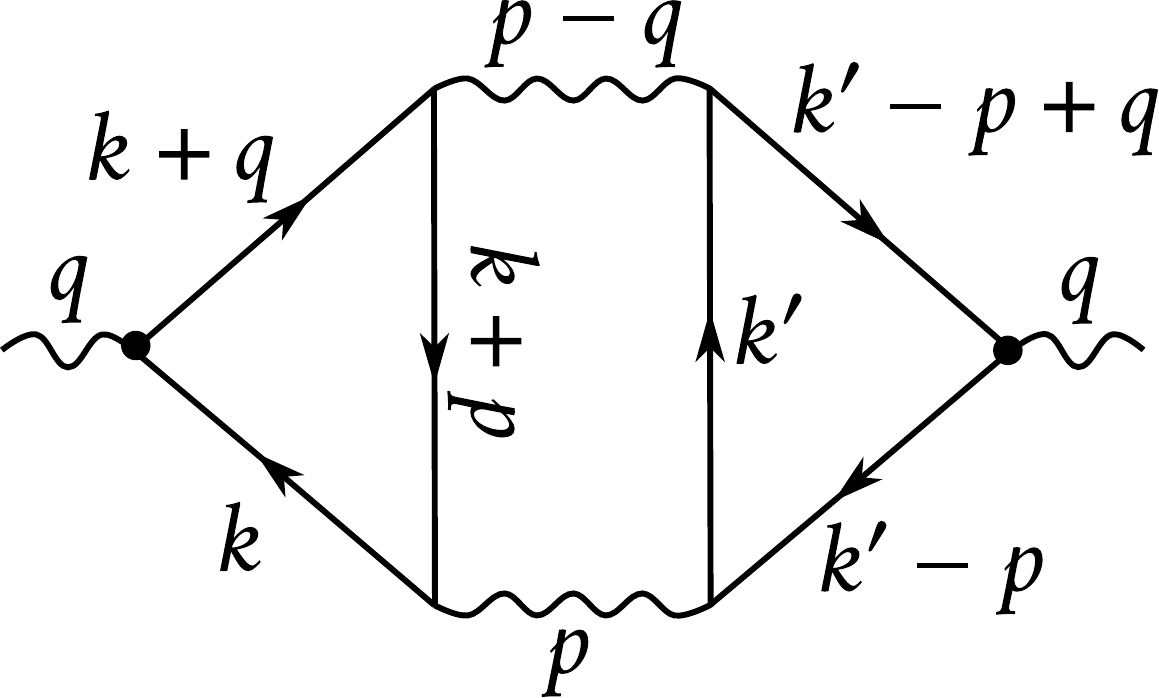}
	 \caption{
	  \label{fig:AL2}}
	\end{subfigure}
	\RawCaption{\caption{Second order diagrams contributing to the normal conductivity: (a,b) The two self-energy (SE) diagrams, (c) the Maki-Thompson (MT) diagram, (d,e) the two Aslamazov-Larkin diagrams  $\mathrm{AL_1}$ and $\mathrm{AL_2}$.}
	\label{fig:NC}}
\end{figure}

The correlator $\Pi (\omega)$ is diagrammatically given by a fully dressed particle-hole bubble with fermionic current in the vertices.  For free fermions, $\Pi (\omega)$ is real and does not contribute to conductivity. Renormalizations by fermion-boson interactions make $\Pi (\omega)$ complex. To leading order in the fermion-boson coupling, the dressed $\Pi (\omega)$ contains distinct contributions from: (i) dressing a fermionic line in the bubble by a self-energy (SE) insertion, Figs.~\ref{fig:se1},\ref{fig:se2}, (ii) inserting  Maki-Thompson (MT) type vertex correction, Fig.~\ref{fig:MT}, and (iii-iv) inserting two topologically non-equivalent Aslamazov-Larkin (AL) vertex corrections, Figs.~\ref{fig:AL1},\ref{fig:AL2}. At a QCP, each of these contributions is proportional to $\sigma (\omega) \propto 1/\omega^{4/3}$. However, for scattering by a small angle $\theta$, the sum of SE and MT contributions, as well as the sum of the two AL contributions, is suppressed by a factor of $(1-\cos{\theta}) \approx \theta^2/2$. Because typical $\theta \sim \omega^{1/3}$, one gets $\sigma_{SE+MT} \sim 1/\omega^{2/3}$ and $\sigma_{AL_1+AL_2} \sim 1/\omega^{2/3}$, instead of $1/\omega^{4/3}$. This reduction can be viewed as the consequence of an \textit{approximate} Ward identity as for the small-angle scattering the renormalization of the charge current vertex mirrors that of the \textit{charge density} vertex. For the latter, the  cancellation would be exact by Ward identity associated with charge conservation.

For a Galilean-invariant system, $\sigma_{SE+MT}$ and $\sigma_{AL_1+AL_2}$ cancel each other, in line with the general requirement that conductivity in such a system must vanish at any finite frequency. However, for a non-Galilean-invariant system, there is no symmetry constraint that would relate these two components. It was argued in Ref.~\cite{PhysRevB.50.17917} that in this situation the net conductivity $\sigma \propto 1/\omega^{2/3}$~\footnote{The same $1/\omega^{2/3}$ behavior of the conductivity~\cite{PhysRevB.94.045133} has been argued to hold at a $q=0$ QCP towards loop-current order~\cite{10.21468/SciPostPhys.14.5.113,10.21468/SciPostPhys.13.5.102}.}. However, subsequent studies have found that in 2D, the conductivity may be further reduced by additional geometric restrictions on scattering processes~\cite{PhysRevLett.106.106403}. For a circular FS, the cancellation of the $1/\omega^{2/3}$ contributions to conductivity from $\sigma_{SE+MT}$ and $\sigma_{AL_1+AL_2}$ has been explicitly demonstrated in Refs.~\cite{PhysRevB.106.115151,guo2023fluctuation,10.21468/SciPostPhys.14.5.113}. It was further argued in~\cite{PhysRevLett.106.106403,shi2023controlled}  that these restrictions hold for a circular, and, more generally, convex FS, but not for a concave FS with an inflection point, as for the latter there exist additional scattering channels, in which near-cancellation between $\sigma_{SE+MT}$ and $\sigma_{AL_1+AL_2}$ should not hold, hence $\sigma (\omega)$ should remain $1/\omega^{2/3}$.

The computation of the conductivity at a QCP in the expansion in the fermion-boson coupling is somewhat questionable because for a non-Fermi liquid the self-energy is not a weak perturbation. In a recent paper~\cite{PhysRevB.108.235125} the authors computed the conductivity away from a QCP, when the correlation length $\xi$ for nematic fluctuations is large but finite, and the system preserves a Fermi liquid behavior at the lowest frequencies. They obtained the formulas for $\sigma (\omega, \xi)$ in the Fermi liquid regime at the smallest $\omega$: $\sigma (\omega, \xi) \sim \omega^2 \xi^4 \log{\omega \xi^3}$ for a convex FS and $\sigma(\omega, \xi) \sim \xi^2$  for a concave FS\@. They conjectured that there is a single crossover scale between a Fermi liquid and a non-Fermi liquid at $\omega \sim \xi^{-3}$.  Substituting $\omega^{-1/3}$ instead of $\xi$, they reproduced the expected $\sigma (\omega) \sim \omega^{2/3}$ for a convex FS and $\sigma(\omega) \sim 1/\omega^{2/3}$ for a concave FS\@.

The authors of~\cite{PhysRevB.108.235125} didn't compute $\Pi$ diagrammatically. Instead, they followed Refs.~\cite{PhysRevB.72.104510,https://doi.org/10.1002/andp.200651807-809} and expressed the conductivity via the retarded correlator of the \textit{time derivatives} of the current, $K(t) \equiv \partial_t \bm{j} = i [H, \bm{j}]$, where $H$ is the Hamiltonian, expressed $K(t)$ explicitly as four-fermion operators, and evaluated the retarded correlator of $K$ using Wick's theorem.

In this work, we compute  conductivity in a Fermi liquid regime near a nematic QCP by using the conventional Kubo formula, Eq.~\eqref{eq:a}, relating conductivity to the retarded current-current correlator. We explicitly evaluate diagrammatically the four contributions to the current-current correlator $\Pi$. We largely reproduce the results of Ref.~\cite{PhysRevB.108.235125}, up to one relatively minor discrepancy.

We also analyze  additional term in the conductivity,  associated with the  ``anomalous'' component of the current~\cite{PhysRevB.108.235125}, proportional to the gradient of the $d-$wave form-factor~\footnote{The anomalous contribution to conductivity has been identified also for fermions coupled to $U(1)$ gauge field~\cite{PhysRevB.50.17917}.}. This term is absent in a Galilean-invariant system, but does emerge when Galilean-invariance is broken. The strength of this term depends on the underlying fermion-boson model. In our study, we  consider a  model, relevant to $\mathrm{FeSe}$, in which a nematic order spontaneously breaks the symmetry between $d_{xz}$ and $d_{yz}$ orbitals of $\mathrm{Fe}$~\cite{Fernandes_2017}. This is a manifestly non-Galilean-invariant lattice model, even if fermionic  dispersion can be approximated by a rotationally-invariant, parabolic form. As a consequence of the absence of rotational invariance,  the $d-$wave form-factor $F(\bm{k},\bm{k}',\bm{p})$ in the boson-mediated 4-fermion interaction
\begin{equation}
	H = \sum_{\bm{k},\bm{k}',\bm{p}} F(\bm{k},\bm{k}',\bm{p}) V(\bm{p}) c^\dagger_{\bm{k}+\tfrac{\bm{p}}{2}} c_{\bm{k}-\tfrac{\bm{p}}{2}}^{\vphantom{\dagger}} c^\dagger_{\bm{k}'-\tfrac{\bm{p}}{2}} c_{\bm{k}'+\tfrac{\bm{p}}{2}}^{\vphantom{\dagger}}
\end{equation}
is determined by coherence factors associated with the transformation from orbital to band states, and at small $p$ can be factorized between $\bm{k}$ and $\bm{k}'$ into $f_{\bm{k}} f_{\bm{k}'}$ (see Eq.~\eqref{eq:ham} below). The factor $f_{\bm{k}}$ for $\bm{k}$ on the FS can in turn be well approximated by $f_{\bm{k}} =\cos{2 \theta_{\bm{k}}}$~\cite{PhysRevB.97.155115,islam2023unconventional}, where $\theta_{\bm{k}}$ is the angle between $\bm{k}$ and, e.g., $x$ axis~\footnote{ For a rotationally-invariant system, $F(\bm{k},\bm{k}',0)$ depends on $\theta_{\bm{k}} - \theta_{\bm{k}'}$ and cannot be factorized into a single product of  $f_{\bm{k}} f_{\bm{k}'}$.}. For such a system, we identify two contributions to conductivity: the ``normal'' one, which contains powers of $f^2_{\bm{k}}$, and the ``anomalous'' one, which contains ${(\grad_{\bm{k}} f_{\bm{k}})}^2$. The ``normal'' contribution comes predominantly from momenta for which $\cos^2(2\theta_{\bm{k}}) \approx 1$, i.e., in this contribution $f_{\bm{k}}$ can be safely approximated by one. For ``anomalous'' contribution, the angular dependence of $f_{\bm{k}}$ is crucial.

We compute the conductivity to second order in fermion-boson coupling $g$, but go beyond this order to ascertain the correct argument under the logarithm (see below). For non-parabolic dispersion, we find that the normal contribution supersedes the anomalous one, yielding $\sigma (\omega) \propto \omega^2 \xi^4 \log{\omega \xi^3}$ for a convex FS, and $\sigma (\omega)  \propto \xi^2$ for a concave FS\@. These results are in agreement with Ref.~\cite{PhysRevB.108.235125}. For a circular FS and parabolic dispersion, we find that the ``normal'' contribution to conductivity vanishes, but the  anomalous contribution remains non-zero. We find that the conductivity in this case is $\sigma (\omega) \propto \omega^2 \xi^4$. The authors of Ref.~\cite{PhysRevB.108.235125} obtained the same form, but with extra $\log{\omega \xi^3}$.

\begin{figure}[htb]
	\centering
	  \includegraphics[width=0.45\linewidth]{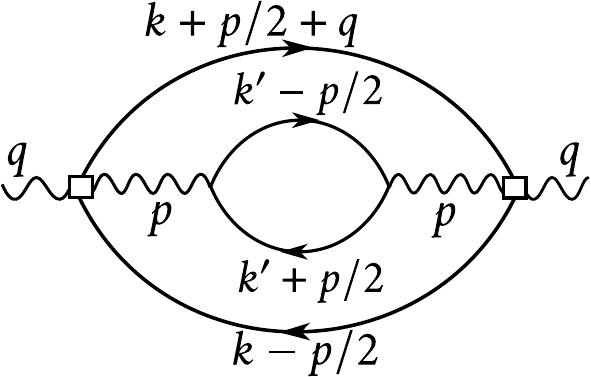}
	\caption{The second-order contribution to conductivity  from the anomalous current — the one with $\grad_{\bm{k}} f_{\bm{k}}$ in the external vertices (squares).
	\label{fig:AN}}
\end{figure}

\section{The model}

We consider a fermion-boson system near a nematic QCP, described by
\begin{equation}
    \label{eq:ham}
        H = \sum_{\bm{k}} \varepsilon_{\bm{k}} c_{\bm{k}}^{\dagger} c_{\bm{k}}^{\vphantom{\dagger}} + \sum_{\bm{p}}
        \omega_{\bm{p}} b_{\bm{p}}^{\dagger} b_{\bm{p}}^{\vphantom{\dagger}} + g \sum_{\bm{k} \bm{p}} f_{\bm{k}} (b_{\bm{p}}^{\vphantom{\dagger}} +  b_{-\bm{p}}^{\dagger}) c_{\bm{k} + \frac{\bm{p}}{2}}^{\dagger}
        c_{\bm{k} - \frac{\bm{p}}{2}}^{\vphantom{\dagger}}\,.
\end{equation}
Here $c_{\bm{k}}$ and $b_{\bm{p}}$ are the electron and boson annihilation operator, respectively, $\varepsilon_{\bm{k}}$ and $\omega_{\bm{p}}$ are free electron and boson dispersions, $g$ is a coupling constant, and a $d-$wave form-factor $f_{\bm{k}} = \cos(2 \theta_{\bm{k}})$.

The free electron and boson propagators are, correspondingly,
\begin{equation}
	\label{eq:el_propagator}
	G(k) = \frac{1}{i k_0 - (\varepsilon_{\bm{k}} - \mu)}
\end{equation}
and
\begin{equation}
	\label{eq:bos_propagator}
     D(\bm{p}) = \frac{\chi_0}{\bm{p}^2 + \xi^{-2}}\,.
\end{equation}

Here $k \equiv (k_0,\bm{k})$ is a 1+2 component vector with a 1D frequency $k_0$ and 2D momentum $\bm{k}$ components, $\mu$ is the chemical potential, and $\xi$ is the correlation length. It is convenient to introduce the effective electron-boson coupling $\bar{g} = g^2 \chi_0$, which has the dimension of energy.

The optical conductivity is determined by the Kubo correlator of the longitudinal components of the current operator. The paramagnetic part of the current is obtained from the variation of the Hamiltonian with the minimal coupling $\bm{k} \to \bm{k} + \bm{A}$ over the vector potential. This yields
\begin{equation}
	\bm{J} \equiv \frac{\delta H}{\delta \bm{A}} {\bigg|}_{A=0} =  \bm{J}_{N} + \bm{J}_{A}\,,
\end{equation}
where the normal component $\bm{J}_{N}$ comes from the electron dispersion,
\begin{equation}
	\label{eq:current_normal}
    \bm{J}_{N} = \sum_{\bm{k}} \bm{v}_{\bm{k}} c_{\bm{k}}^{\dagger} c_{\bm{k}}^{\vphantom{\dagger}}\,,
\end{equation}
with the group velocity $\bm{v}_{\bm{k}} \equiv \grad_{\bm{k}} \varepsilon_{\bm{k}}$, while the anomalous component
\begin{equation}
	\label{eq:current_anomalous}
    \bm{J}_{A} = g \sum_{\bm{k} \bm{p}} \grad_{\bm{k}} f_{\bm{k}} (b_{\bm{p}}^{\vphantom{\dagger}} +
    b_{-\bm{p}}^{\dagger}) c_{\bm{k} + \frac{\bm{p}}{2}}^{\dagger} c_{\bm{k} - \frac{\bm{p}}{2}}^{\vphantom{\dagger}}
\end{equation}
comes from the momentum (angular) dependence of electron-boson interaction. The gradient of the $d-$wave form-factor
in polar coordinates has the form of $\grad_{\bm{k}} f_{\bm{k}} = 2 k^{-1}_{F} \sin (2 \theta_{\bm{k}})\, \bm{e}_{\theta}$.

We note that both components of the current must be kept to satisfy the continuity equation for electron density. Indeed, the normal component $\bm{J}_N$ of the total current compensates for the density commutator with the free electron dispersion. The electron density commutator with the interaction part of the Hamiltonian, given by the last
term of Eq.~\eqref{eq:ham}, equals
\begin{equation}
    \dot{\rho}_{\bm{q}} = i [H_{\mathrm{int}},\rho_{\bm{q}}]
	=  i g \sum_{\bm{k} \bm{p}} (f_{\bm{k} + \bm{q}} - f_{\bm{k}}) (b_{\bm{p}}^{\vphantom{\dagger}} +
b_{-\bm{p}}^{\dagger}) c_{\bm{k} + \bm{q} + \frac{\bm{p}}{2}}^{\dagger} c_{\bm{k} -
\frac{\bm{p}}{2}}^{\vphantom{\dagger}}\,,
\end{equation}
In the long-wave limit of $\bm{q} \to 0$, it is  compensated by the anomalous current of Eq.~\eqref{eq:current_anomalous}.\\

\section{Conductivity}

The diagrammatic computation of the current-current correlator is rather straightforward. We present the details in Appendix~\ref{app_pol} and here quote the results. The contribution to $\Pi (\omega)$ from the normal component of the current (Fig.~\ref{fig:NC}) is
\begin{widetext}
	\begin{equation}
		\label{eq:polarization}
		\begin{split}
			\Pi (\omega) =& \frac{\pi^3}{2 \omega^2} \frac{\bar{g}^2}{\chi_0^2} \int \frac{d^2 \bm{k} d^2 \bm{k'}
			d^2 \bm{p}}{{(2\pi)}^9} D^2(\bm{p})  f^2_{\bm{k}} f^2_{\bm{k}'} \frac{{(\Delta \bm{v})}^2}{\omega - \Delta \varepsilon -i0}\\
			&{}\times \left(\sign \varepsilon_{\bm{k} - \frac{\bm{p}}{2}} - \sign \varepsilon_{\bm{k} +
			\frac{\bm{p}}{2}}\right) \left(\sign \varepsilon_{\bm{k}' - \frac{\bm{p}}{2}} - \sign \varepsilon_{\bm{k}' +
			\frac{\bm{p}}{2}}\right) \left(\sign (\varepsilon_{\bm{k} - \frac{\bm{p}}{2}} - \varepsilon_{\bm{k} +
			\frac{\bm{p}}{2}}) - \sign (\varepsilon_{\bm{k}' - \frac{\bm{p}}{2}} - \varepsilon_{\bm{k}' +
			\frac{\bm{p}}{2}})\right)\,.
		\end{split}
	\end{equation}
\end{widetext}
Here
\begin{equation}
	\label{eq:velocity_change}
	\Delta \bm{v} \equiv \bm{v}_{\bm{k}'-\frac{\bm{p}}{2}} + \bm{v}_{\bm{k}+\frac{\bm{p}}{2}} -
	\bm{v}_{\bm{k}'+\frac{\bm{p}}{2}} - \bm{v}_{\bm{k}-\frac{\bm{p}}{2}}
\end{equation}
is the velocity change, and
\begin{equation}
	\label{eq:energy_change}
	\Delta \varepsilon \equiv \varepsilon_{\bm{k}'-\frac{\bm{p}}{2}} + \varepsilon_{\bm{k}+\frac{\bm{p}}{2}} -
	\varepsilon_{\bm{k}'+\frac{\bm{p}}{2}} - \varepsilon_{\bm{k}-\frac{\bm{p}}{2}}
\end{equation}
is the energy change in the two-electron scattering process. The $\im \Pi (\omega)$, which we need for conductivity, is obtained by replacing $1/(\omega - \Delta \varepsilon -i0)$ by $i \pi \delta (\omega - \Delta \varepsilon)$.

\begin{figure}[htb]
	\centering
	\begin{subfigure}{.33\linewidth}
	  \centering
	  \includegraphics[width=.9\linewidth]{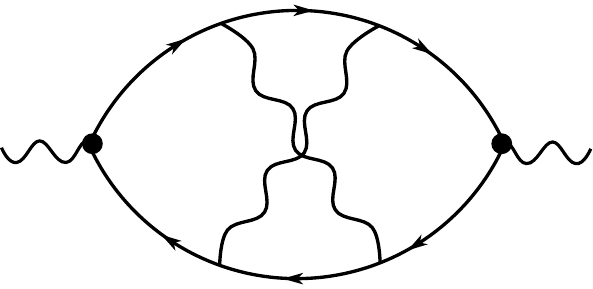}
	  \caption{
	  \label{fig:mt_crossed}}
	\end{subfigure}%
	\centering
	\begin{subfigure}{.33\linewidth}
	  \centering
	  \includegraphics[width=.9\linewidth]{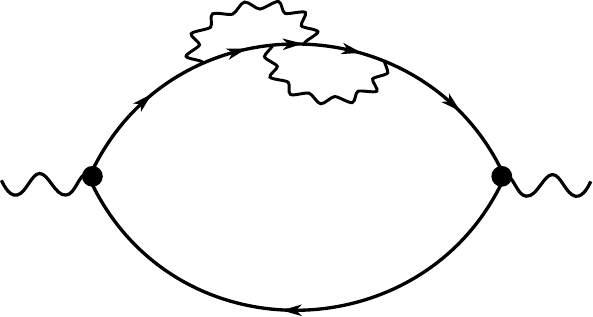}
	  \caption{
	  \label{fig:se_crossed}}
	\end{subfigure}%
	\begin{subfigure}{.33\linewidth}
	  \centering
	  \includegraphics[width=.9\linewidth]{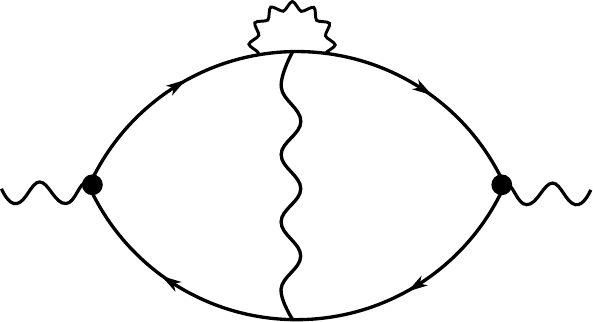}
	  \caption{
	  \label{fig:mt_vertex}}
	\end{subfigure}
	\RawCaption{\caption{Other contributions to the current current correlator with two bosonic propagators. These contributions are small  in $1/(k_F \xi)$ compared to the ones we included in our analysis. \label{fig:other}}}
\end{figure}

For completeness, we  also analyzed other diagrams for the current-current correlator with two bosonic propagators. We show these additional diagrams in Fig.~\ref{fig:other}. We found that their contribution is suppressed by a factor of $1/(k_F \xi)$ compared to Eq.~\eqref{eq:polarization}, because the two bosonic propagators in the diagrams in Fig.~\ref{fig:other} are with different momenta, and the integration over momentum difference yields a factor $1/\xi$ instead of $k_F$ in  Eq.~\eqref{eq:polarization}. 

The contribution to $\Pi(\omega)$ from the anomalous current (Fig.~\ref{fig:AN}) can be expresses as a convolution of two fermionic bubbles,
\begin{equation}
	\label{eq:anom_pol}
	\Pi_{A}(\omega) = \frac{\bar{g}^2}{\chi_0^2} \int \frac{d^3 p}{{(2\pi)}^3} D^2(\bm{p}) \Pi_{d}(p) {\bar{\Pi}}_{d}(p,q)\,,
\end{equation}
where $\Pi_{d}(p)$ is the conventional $d-$wave polarization bubble with form-factor $f^2_{\bm{k}}$:
\begin{equation}
	\Pi_{d}(p) = \int \frac{d^3 k}{{(2\pi)}^3} f^2_{\bm{k}} G(k+\tfrac{p}{2})G(k-\tfrac{p}{2})\,,
\end{equation}
and ${\bar{\Pi}}_{d}(p,q)$ is the $d-$wave polarization bubble with form-factor  ${(\grad_{\bm{k}} f_{\bm{k}})}^2$ and frequency shifted by $\omega$ (we denote $q \equiv (\omega, 0)$):
\begin{equation}
	{\bar{\Pi}}_{d}(p,q) = \int \frac{d^3 k}{{(2\pi)}^3} {(\grad_{\bm{k}} f_{\bm{k}})}^2 G(k+\tfrac{p}{2}+q)G(k-\tfrac{p}{2}) \,.
\end{equation}
There are other contributions to $\Pi (\omega)$ from anomalous current, but we verified that they are smaller in $1/(k_F \xi)$. Also, the cross-contribution to $\Pi$ from correlator of $\bm{J}_N$ and $\bm{J}_A$ vanishes because of orthogonality of $\bm{v}_{\bm{k}}$ and $\grad_{\bm{k}} f_{\bm{k}}$.

\subsection{Systems with parabolic dispersion}

We start with  the case when fermionic dispersion $\epsilon_k$ is rotationally invariant and can be approximated as parabolic:  $\varepsilon_k = k^2/2m$. We remind that the fermion-boson Hamiltonian is manifestly non-Galilean-invariant. For rotationally-invariant dispersion, the angle $\theta_k$ in the form-factor $f_{\bm{k}}$ can be measured from any direction. It is convenient to measure it from the direction of the bosonic momentum ${\bm p}$.

Since for quadratic dispersion the velocity $\bm{v}_{\bm{k}}$ is linear in momentum $\bm{k}$, the velocity change of Eq.~\eqref{eq:velocity_change} vanishes, i.e., there is no contribution to conductivity from the normal current, hence $\Pi (\omega) = \Pi_A (\omega)$.

The anomalous contribution can be most straightforwardly evaluated in Matsubara frequencies, for $p = (i p_{0,m},p)$ and $q = (i\omega_m, 0)$. The result  for real $\omega$ is then obtained by replacing $i\omega_m$ by $\omega + i 0$. We assume and verify that $\Pi_A (i\omega_m)$ is determined by internal $p_{0,m} \sim v_F p \sim \omega_m$. At small $\omega_m$, which we are interested in, one can then compute the polarization bubbles in the long-wavelength limit. On the Matsubara axis we have
\begin{equation}
	\Pi_{d}(p) = - \frac{m}{4\pi} \big(1 -\frac{2 \abs{\zeta}}{\sqrt{1+\zeta^2}} + 4\zeta^2 +8\zeta^4 -8 \abs{\zeta}^3 \sqrt{1+ \zeta^2} \big)\,,
\end{equation}
and
\begin{equation}
	\begin{split}
		{\bar \Pi}_{d}(p,q) = &{}- \frac{m}{\pi k_F^2} \Big( 1- 4{(\zeta+\eta)}^2 -8{(\zeta+\eta)}^4\\
		&{} +8 \abs{\zeta+\eta}^3 \sqrt{1+ {(\zeta+\eta)}^2} \Big)\,,
	\end{split}
\end{equation}
where  $\zeta \equiv \frac{m p_{0,m}}{k_F p}$ and $\eta = \frac{m \omega_m}{k_F p}$. In these variables, $\int d^3 p \to \int p^2 dp \int d \zeta$.  The $\abs{\zeta}$ term in $\Pi_{d}(p)$ is a conventional Landau damping. For ${\bar \Pi}_{d}(p,q)$, there is no Landau damping, and the expansion in $(\zeta + \eta)$ starts with the quadratic term. The absence of Landau damping is the consequence of ${(\sin{2\theta_k})}^2$ form of the form factor: Landau damping comes from $\theta_k \approx \pm \pi/2$, for which this form-factor vanishes. To understand the form of $\Pi_A (q)$, we expand the integrand in $\Pi_A (q)$ in $\eta$. The expansion holds in even powers of $\eta$, i.e., in even powers of $\omega_m$. The prefactor for $\eta^2$ is regular in the infra-red and  determined by $\zeta = O(1)$ and $p \sim \xi^{-1}$. This term, however, does not give rise to $\im \Pi (\omega)$ and hence is irrelevant to conductivity. Subtracting this term, we find that by power counting, $\Pi (\omega_m) \propto \omega^3_m$, and typical $p_0$ and $v_F p$ are of order $\omega_m$. On the real frequency axis, this contribution is imaginary and hence relevant to conductivity. At small frequencies, when  $\omega \ll v_F \xi^{-1}$, we can then approximate $D({\bm p})$ in~\eqref{eq:anom_pol} by $D(0)$. Evaluating the integral over $p$ and $\zeta$ and converting to real $\omega$, we obtain
\begin{equation}
	\label{eq:anom_sigma}
		\sigma (\omega) = \sigma_{A}(\omega) = 0.888 e^2 \bar{g}^2 {\left(\frac{\nu}{4 \pi}\right)}^2
		{\left(\frac{\omega}{\varepsilon_F}\right)}^2 \xi^4,
\end{equation}
where $\nu =\frac{m}{2 \pi}$ the density of states and $\varepsilon_F$ the Fermi energy. To high numerical accuracy, the prefactor is $8/9$.  We note here that infra-red convergence of the integral that gives the prefactor is the consequence of the absence of Landau damping term in ${\bar \Pi}_{d}(p,q)$. If the linear in frequency term were present in ${\bar \Pi}_{d}(p,q)$, we would get an additional logarithmic factor $\log{(\omega \xi^3)}$.

We see that for a parabolic dispersion, $\sigma (\omega) \propto \omega^2 \xi^4$. Using $\xi\sim 1/\omega^{1/3}$ for the crossover to the quantum-critical regime, we find that at a QCP, the conductivity behaves $\sigma (\omega) \sim \omega^{2/3}$. This agrees with earlier estimates (see above), but we emphasize that this conductivity comes exclusively from the anomalous component of the current.

\subsection{Systems with non-parabolic isotropic dispersion}

\begin{figure}[htb]
	\centering
	\begin{subfigure}{.5\linewidth}
	  \centering
	  \includegraphics[width=\linewidth]{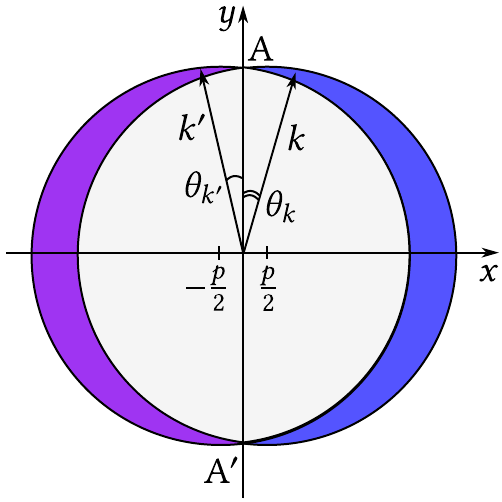}
	  \caption{
	  \label{fig:cnv1}}
	\end{subfigure}%
	\begin{subfigure}{.5\linewidth}
	  \centering
	  \includegraphics[width=\linewidth]{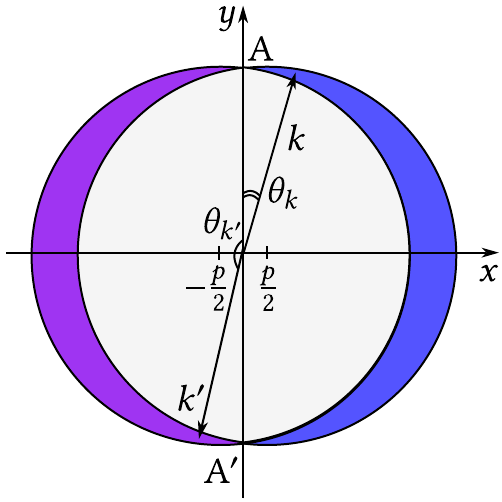}
	  \caption{
	  \label{fig:cnv2}}
	\end{subfigure}
	\caption{The swap (a) and Cooper (b) scattering channels for a circular FS\@. The same holds for any convex FS\@. The circles are the two FSs, shifted by $p$, which we directed along the $x$ axis.
	\label{fig:convex}}
\end{figure}

We next analyze the conductivity for the case when the dispersion $\epsilon_k$ can still be approximated as isotropic, but is not parabolic. To be specific, we consider the dispersion of the form
\begin{equation}
	\varepsilon_{\bm{k}} = \frac{\bm{k}^2}{2 m} + \lambda \frac{k^4}{4}\,.
\end{equation}

Consider first the contribution from the normal current. For a non-zero $\lambda$, the factor $\Delta \bm{v}$ in~\eqref{eq:velocity_change} is also non-zero and for $k$ and $k'$ on the FS is given by
\begin{equation}
	\Delta \bm{v} = 2\lambda \left((\bm{k}\cdot \bm{p}) \bm{k} - (\bm{k}'\cdot \bm{p}) \bm{k}' \right)\,.
\end{equation}

One can easily verify that the integration over the 2D electron momentum is confined to the range in between the two Fermi seas, displaced relative to each other by the bosonic momentum $\bm{p}$. This region is composed of several disjoint domains, with their count corresponding to the number of intersection points between the shifted FSs (see Fig.~\ref{fig:convex}).

In the $\bm{k}$-space, the intersection points are determined by solving the equation $\varepsilon_{\bm{k}-\bm{p}/2} = \varepsilon_{\bm{k}+\bm{p}/2}$. Because electron dispersion has inversion symmetry, intersection points always occur in pairs at ${\bm k}$ and $-{\bm k}$. A pair of any two intersection points defines a \textit{channel of electron-electron scattering}. The contributions from different scattering channels add together in $\Pi (\omega)$. The regions around the intersection points are critically important for conductivity because when electrons are near these points, $\Delta \varepsilon$ is small and can be made equal to $\omega$.

Specifically, for a circular FS, the integration domains are crescent-shaped, as depicted in Fig.~\ref{fig:convex}, with intersections labeled as $A$ and $A'$. Each domain is assigned an integer index, $\pm 1$, with adjacent domains alternating in sign. The factor in the second line in Eq.~\eqref{eq:polarization} ensures that $\bm{k}$ and $\bm{k}'$ are situated in domains with opposing indices; in Fig.~\ref{fig:convex}, these are indicated by distinct colors.

The energy change is given by
\begin{equation}
	\label{eq:energy_change_expanded}
	\Delta \varepsilon = v_F \bm{p} \cdot (\hat{k} - \hat{k}') = v_F p (\sin \theta_k + \sin \theta_{k'}) \,,
\end{equation}
where $v_F = v_{k=k_F}$ and $\hat{k} \equiv \bm{k}/k$, with angles $\theta_k$ and $\theta_{k'}$, which here is convenient to measure from the direction normal to $\bm{p}$. Both angles are in the interval $(0,\pi)$, such that $\sin \theta_{k}$ and $\sin \theta_{k'}$ are positive.

Both $\sin \theta_{k}$ and $\sin \theta_{k'}$ must be small to satisfy $\Delta \varepsilon = \omega$. This condition selects two distinct scattering channels. The first, known as the swap channel, involves electron scattering near the same intersection point, such that $\bm{k} \approx \bm{k}'$ ($\theta_k \approx \theta_{k'} \ll 1$). The second, the Cooper channel, corresponds to the scattering of electrons near points that are related by inversion symmetry, i.e., where $\bm{k} \approx -\bm{k}'$ ($\theta_{k'} \approx \pi - \theta_k, \theta_k \ll 1$). These scattering processes are depicted in Fig.\ref{fig:cnv1} and Fig.\ref{fig:cnv2}, respectively. In both cases, the integration is confined to the vicinity of the intersection points, where $k$ varies within the range $k_F \pm p \theta_k/2$.

Performing integration over $k$ and $k'$ and combining the contributions from the Cooper and swap channels we obtain
\begin{widetext}
	\begin{equation}
			\im \Pi (\omega) = \frac{64 \lambda^2 \pi^4}{\omega^2} \frac{\bar{g}^2}{\chi_0^2} \int \frac{d^2
\bm{p}}{{(2\pi)}^9} D^2(\bm{p}) p^4 \int d\theta_k d\theta_{k'} f^2_{\theta_k} f^2_{\theta_{k'}} \left( \theta_k^3
\theta_{k'} + \theta_k \theta_{k'}^3\right) \delta(\omega - v_F p (\theta_k +  \theta_k'))\,.
	\end{equation}
\end{widetext}
Performing angular integration, we arrive at
\begin{equation}
	\label{eq:apost}
		\im \Pi (\omega) = \frac{\lambda^2 \omega^3}{40 \pi^4} \frac{\bar{g}^2}{v_F^6\chi_0^2} \int_{\Lambda}^{\infty} dp\,\frac{D^2(p)}{p}\,.
\end{equation}
The momentum integral is logarithmically singular and requires an infrared cutoff, $\Lambda$, which we determine below. The ultra-violet cutoff for the logarithmic behavior is $p \sim \xi^{-1}$, set by $D(p)$.  Setting the limits,  we obtain
\begin{equation}
	\label{eq:convex_sigma}
	\sigma(\omega) = \frac{\lambda^2 e^2 \bar{g}^2}{40 \pi^4 v_F^6} \omega^2 \xi^4 \log \xi \Lambda \,.
\end{equation}
Strictly within perturbation theory, where $D(p)$ is static, the infra-red cutoff is set at $L \sim \omega/v_F$. However, near a QCP, the dynamics of $D(p)$ (the Landau damping) cannot be neglected. Using the full dynamical form of the bosonic propagator along the Matsubara axis,
\begin{equation}
	\label{eq:int_bos_propagator}
     \mathcal{D}(p) = \mathcal{D}(\bm{p},\omega_p) = \frac{\chi_0}{\bm{p}^2 + \xi^{-2}+ \gamma \tfrac{\abs{\omega_p}}{p}}\,,
\end{equation}
with $\gamma \propto {\bar g}$, we find that once $\gamma v_F \xi^2 >1$, the lower cutoff is set by $\Lambda \sim \gamma \xi^2 \omega$. In this last regime, the normal component of conductivity scales as $\sigma_N (\omega) \sim \omega^2 \xi^4 \log (\gamma \omega \xi^3)$. The contribution from the anomalous current does not change qualitatively when $\lambda$ is finite and the anomalous component of conductivity remains $\sigma_A (\omega) \sim \omega^2 \xi^4$. We see that in the Fermi liquid regime, where $\omega$ is small and $\log (\gamma \omega
\xi^3)$ is large, the normal component of $\sigma$ is logarithmically larger than the anomalous component.

As before, we extrapolate Eq.~\eqref{eq:convex_sigma} to the nematic QCP  by replacing $\xi \sim \omega^{-1/3}$. The conductivity becomes $\sigma(\omega) \sim \omega^{\frac23}$. Note that the argument of the logarithm becomes of order one.

For completeness, we also evaluated the contributions to conductivity separately from SE, MT, an $\mathrm{AL}_1$ and $\mathrm{AL}_2$ diagrams for the correlator of the normal current. We found that each individual  contribution to conductivity scales as $\xi^4 \log{(\omega \xi^3)}$. At a nematic QCP this becomes $1/\omega^{4/3}$. We see that the total contribution is far smaller that individual ones. At a QCP, the smallness is an extra $\omega^2$ factor in the full $\sigma (\omega)$.

\subsection{Convex FS}

The scaling form of the conductivity in Eq.~\eqref{eq:convex_sigma}  holds also a more general case of a non-isotropic dispersion, when the FS has a convex form. Once can verify~\cite{PhysRevLett.106.106403} that  any convex FS in a 2D system, the intersections of the shifted FSs are limited to no more than two points, the analogs of $A$ and $A'$ in Fig.~\ref{fig:convex}. This constrains the potential scattering channels to only Cooper and swap ones. As the consequence, the leading contributions from  SE+MT and $\mathrm{AL}_1+\mathrm{AL}_2$ diagrams still cancel out \textit{within each scattering channel}. The leftover conductivity is still given by Eq.~\eqref{eq:convex_sigma}. In practical terms, the SE+ MT and $\mathrm{AL}_1+\mathrm{AL}_2$, taken separately, yield $\im \Pi (\omega)$ as in Eq.~\eqref{eq:polarization}, but with ${(\Delta \bm{v})}^2$ replaced by $\bm{v}_{\bm{k}-\frac{\bm{p}}{2}} \cdot (\bm{v}_{\bm{k}+\frac{\bm{p}}{2}} - \bm{v}_{\bm{k}-\frac{\bm{p}}{2}})$ for SE +MT diagrams and $\bm{v}_{\bm{k}-\frac{\bm{p}}{2}} \cdot (\bm{v}_{\bm{k}'-\frac{\bm{p}}{2}} -
\bm{v}_{\bm{k}'+\frac{\bm{p}}{2}})$ for $\mathrm{AL}_1+\mathrm{AL}_2$ diagrams, see Eqs.~(\ref{Eq:app-ref-1}--\ref{Eq:app-ref-3}) in the Appendix.

Each contribution yields  $\im \Pi(\omega) \sim \omega \bar{g}^2 \xi^2$ with the same prefactor, but the signs are opposite. This becomes evident when we expand the velocity combinations to the second order in the bosonic momentum:
\begin{equation}
	\label{Eq:MT_vel_expansion}
		\bm{v}_{\bm{k}-\frac{\bm{p}}{2}} \cdot (\bm{v}_{\bm{k}+\frac{\bm{p}}{2}} - \bm{v}_{\bm{k}-\frac{\bm{p}}{2}}) \sim
		-\frac12 p_{\beta} p_{\gamma} \frac{\partial v_{\alpha}}{\partial k_{\beta}} \frac{\partial v_{\alpha}}{\partial
		k_{\gamma}} \,,
\end{equation}
while
\begin{equation}
	\label{Eq:AL_vel_expansion}
		\bm{v}_{\bm{k}-\frac{\bm{p}}{2}} \cdot (\bm{v}_{\bm{k}'-\frac{\bm{p}}{2}} - \bm{v}_{\bm{k}'+\frac{\bm{p}}{2}})
		\sim \frac12 p_{\beta} p_{\gamma} \frac{\partial v_{\alpha}}{\partial k_{\beta}} \frac{\partial v_{\alpha}}{\partial
		k'_{\gamma}}\,.
\end{equation}
Evaluating the sub-leading terms, which do not cancel between SE+MT and  $\mathrm{AL}_1+\mathrm{AL}_2$, we obtain $\im \Pi (\omega) \sim \omega^3 \xi^4 \log{\omega \xi^3}$ and $\sigma (\omega) \sim \omega^2 \xi^4 \log{\omega \xi^3}$, like in  Eq.~\eqref{eq:convex_sigma}.

\subsection{Concave FS}

\begin{figure}[htb]
	\centering
	\includegraphics[width=.5\linewidth]{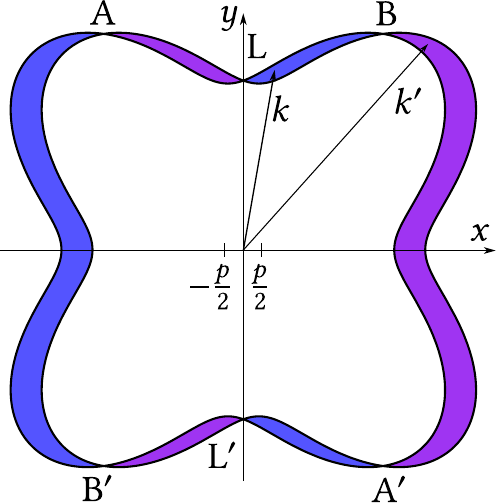}
	\caption{Additional scattering channel for a concave FS\@. The colored region is between the two FSs, shifted by $p$, which we directed along the $x$ axis.
	A, B, and L are crossing points between the FSs. Points A and B exist only for a concave FS\@. The momenta $\bm{k}$ and $\bm{k}'$ belong to topologically different domains, which we depicted by distinct colors.
	\label{fig:concave}}
\end{figure}

The situation changes qualitatively for a concave FS, see Fig.~\ref{fig:concave}. For such a FS, additional intersection points emerge~\cite{PhysRevLett.106.106403}, as we show in Fig.~\ref{fig:concave} for  bosonic momentum ${\bm p}$ directed along $x$. This gives rise to additional scattering channels, beyond swap and Cooper ones. One new scattering channel showcased in Fig.~\ref{fig:concave} involves electrons with momenta ${\bm k}$  and ${\bm k}'$ near distinct and non-equivalent points $L$ and $B$ (${\bm k}$ near $L$ and ${\bm k}'$ near $B$). For these electrons,  $\frac{\partial v_{\alpha}}{\partial k'_{\gamma}}$ and $-\frac{\partial v_{\alpha}}{\partial k_{\gamma}}$ are not related and, as a result, there is no cancellation between SE+MT and $\mathrm{AL}_1+\mathrm{AL}_2$ contributions.

For a ``weakly'' concave FS, when the inflection points are close to each other, electron dispersion near $L$ and $B$  can be approximated by  $\varepsilon_{\bm{k}} = k_y^2 -2 \delta k_x^2 +k_x^4$, where $\delta$ is small but positive. The point $L$ is at $k_x=0$ and the point $B$ is at $k_x = \frac12 \sqrt{4 \delta - p^2}$.  Evaluating SE+MT and $\mathrm{AL}_1+\mathrm{AL}_2$ contributions in the vicinity of points $L$ and $B$, we find that the leading terms  in SE+MT and $\mathrm{AL}_1+\mathrm{AL}_2$ do not cancel at a finite $\delta$. The sum of SE+MT and $\mathrm{AL}_1+\mathrm{AL}_2$ is still smaller than SE+MT and AL$_1$+AL$_2$ taken separately, but the smallness is by an overall factor  $\delta^{7/2}$. Substituting into Kubo formula, we obtain $\sigma'(\omega) \propto \bar{g}^2 \xi^2 \delta^{\frac72} \vartheta(\delta)$.

Combining this with the earlier results, we find that for a concave FS at small $\delta$, the full optical conductivity behaves as
\begin{equation}
	\sigma(\omega) \sim \bar{g}^2  \left[C_1 \omega^{2} \xi^4 \log{\omega \xi^3} + C_2 \xi^2 \delta^{\frac72}
	\vartheta(\delta)\right] \,.
\end{equation}
where $C_1$ and $C_2$ are the combinations of system parameters. For larger $\delta$, the second term obviously becomes much larger than the first one.

 At a QCP, this becomes
\begin{equation}
	\sigma(\omega) \sim  \left[{\bar C}_1 \omega^{2/3} + \frac{{\bar C}_2}{\omega^{2/3}} \delta^{\frac72}
	\vartheta (\delta) \right]\,.
\end{equation}

\section{Conclusion}
In this study, we analyzed the optical conductivity of a pristine two-dimensional electron gas near the Ising-nematic QCP at zero temperature. We considered the model, in which Galilean invariance is explicitly broken by the lattice and, as a result, the $d-$wave nematic form-factor $F(\bm{k},\bm{k}')$ in the effective 4-fermion interaction $\sum_{\bm{k},\bm{k}',\bm{q}} F(\bm{k},\bm{k}') V(\bm{q}) c^\dagger_{\bm{k}+\bm{q}/2} c_{\bm{k}-\bm{q}/2} c^\dagger_{\bm{k}'-\bm{q}/2} c_{\bm{k}'+\bm{q}/2}$, mediated by soft nematic fluctuations with small $\bm{q}$, can be factorized into $F(\bm{k},\bm{k}') = f_{\bm{k}} f_{\bm{k}'}$. Fermionic dispersion, on the other hand, can still be, in some cases, well described by a parabola.  We considered four cases: (i) a parabolic dispersion, (ii) a non-parabolic, but still isotropic dispersion, (iii) a non-isotropic dispersion and a convex FS, and (iv) a concave FS\@. We computed optical conductivity at the lowest frequencies in the Fermi liquid regime away from the nematic QCP and then extended the results to the QCP by using the scaling between frequency and nematic correlation length $\xi$ ($\xi \to \omega^{-1/3}$). We used Kubo formula, which  relates conductivity to the imaginary part of the fully renormalized current-current polarization bubble $\Pi (\omega)$ and obtained $\Pi$ diagrammatically, to the second order in fermion-boson coupling, but combining contributions from inserting fermionic self-energy and MT and AL-type vertex corrections. We also computed additional contribution to $\Pi (\omega)$ from the anomalous current, proportional to the gradient of the $d-$wave form-factor. In the case (i) we found that optical conductivity comes entirely from the anomalous current and behaves as  $\sigma (\omega) \sim \omega^2 \xi^4$. In cases (ii) and (iii), the conductivity predominantly comes from the normal current and behaves as $\sigma (\omega) \sim \omega^2 \xi^4 \log{\omega \xi^3}$. This conductivity is far smaller than the contribution from each diagram $(\xi^4 \log{\omega \xi^3}$) and even smaller than the combined contribution from SE+MT and from AL$_1$+AL$_2$ ($\xi^2$). The reduction from $\xi^4 \log{\omega \xi^3}$ to $\xi^2$ was expected, since for small-angle scattering the contributions from SE+MT and from AL$_1$+AL$_2$ contain extra factor $(1- \cos{\theta}) \approx \theta^2/2$ as a consequence of near equivalence between current and density vertices for small-angle scattering. Further reduction of conductivity between SE+MT and $\mathrm{AL}_1+\mathrm{AL}_2$ contributions (the cancellation of the leading terms) is specific to 2D and is a consequence of the geometric restrictions imposed on the scattering on a convex FS\@. In case (iv), the reduction by small $(1- \cos{\theta})$ still holds, but there is no geometrical restriction. As a consequence, $\sigma (\omega) \sim \xi^2$. At a QCP, this yields $\sigma (\omega) \sim 1/\omega^{2/3}$ — the same form as has been suggested in the original work by Y-B Kim \textit{et al.}~\cite{PhysRevB.50.17917}.

 \begin{acknowledgments}
	We thank A.~Kamenev, A.~Levchenko, and D.~Maslov for helpful discussions.  The work of Y.G.\ was supported by the Simons Foundation Grant No.~1249376;  A.Ch.\ was supported by U.S. Department of Energy, Office of Science, Basic Energy Sciences, under Award No.~DE-SC0014402. Part of the work was done at the Kavli Institute for Theoretical Physics (KITP) in Santa Barbara, CA\@. KITP is supported by the National Science Foundation under Grants No.~NSF PHY-1748958 and PHY-2309135.
\end{acknowledgments}

\appendix
\section{The derivation of the electron polarization}
\label{app_pol}
\begin{figure}[htb]
	\centering
	\begin{subfigure}{.5\linewidth}
		\centering
		\includegraphics[width=.8\linewidth]{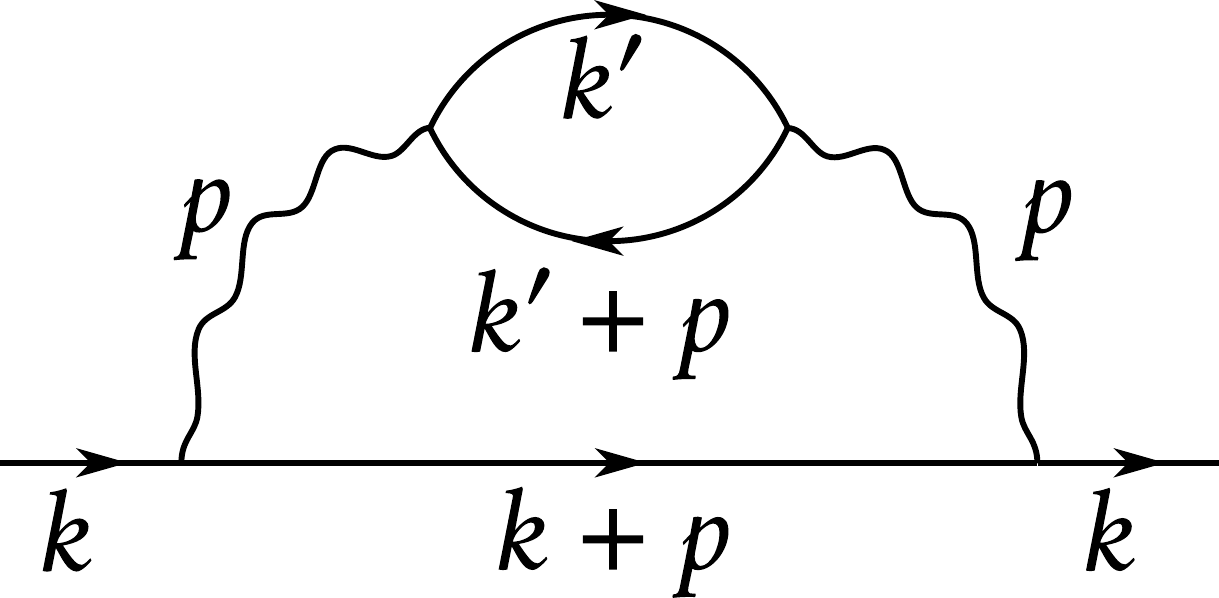}
		\caption{
			\label{fig:sfe}}
		\end{subfigure}%
		\begin{subfigure}{.5\linewidth}
			\centering
			\includegraphics[width=.8\linewidth]{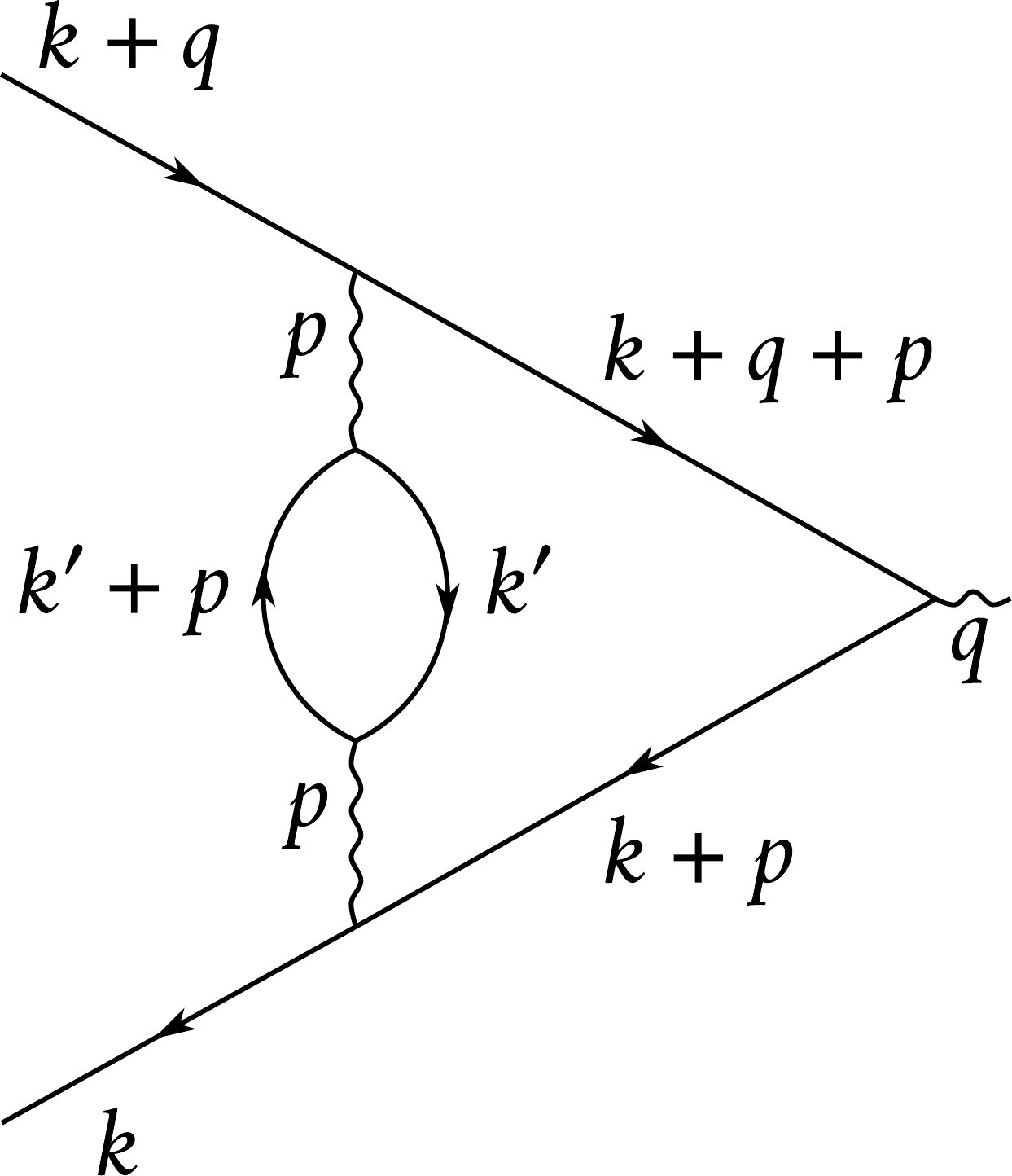}
			\caption{
				\label{fig:sv}}
		\end{subfigure}
		\RawCaption{\caption{(a) The self-energy part. (b) The scalar vertex diagram.}}
\end{figure}
		
\begin{widetext}

Here, we derive the Eq.~\eqref{eq:polarization} for the polarization. To begin with, introduce the self-energy part
given by the diagram in Fig.~\ref{fig:sfe}. It equals
\begin{equation}
    \Sigma(k) = -\frac{\bar{g}^2}{\chi_0^2} \int \frac{d^3 k' d^3 p}{{(2\pi)}^6} D^2(\bm{p})
    f^2_{\bm{k}+\frac{\bm{p}}{2}} f^2_{\bm{k}'+\frac{\bm{p}}{2}} G(k') G(k'+p) G(k+p)\,.
\end{equation}
The scalar vertex shown in Fig.~\ref{fig:sv} is given by
\begin{equation}
    \Gamma(k+q;q) = -\frac{\bar{g}^2}{\chi_0^2} \int \frac{d^3 k' d^3 p}{{(2\pi)}^6} D^2(\bm{p})
    f^2_{\bm{k}+\frac{\bm{p}}{2}} f^2_{\bm{k}'+\frac{\bm{p}}{2}} G(k') G(k'+p) G(k+p) G(k+p+q)\,.
\end{equation}
Since the free electron propagator of Eq.~\eqref{eq:el_propagator} satisfies
\begin{equation}
\label{eq:green_relation}
    G(k+p)G(k+p+q) = \frac{1}{i q_0} \left(G(k+p) - G(k+p+q)\right)\,,
\end{equation}
the scalar vertex is seen to be related to the self-energy part by the Ward identity,
\begin{equation}
     \Gamma(k+q;q) = \frac{\Sigma(k) - \Sigma(k+q)}{i q_0}\,.
\end{equation}

The SE diagrams of Figs.~\ref{fig:se1} and~\ref{fig:se2} are equal, respectively, to
\begin{equation}
    \Pi_{SE_1} = - \int \frac{d^3 k}{{(2 \pi)}^3} G^2(k+q)G(k) \Sigma(k+q) \,\bm{v}_{\bm{k}}^2
\end{equation}
and
\begin{equation}
    \Pi_{SE_2} = - \int \frac{d^3 k}{{(2 \pi)}^3} G(k+q)G^2(k) \Sigma(k) \, \bm{v}_{\bm{k}}^2\,.
\end{equation}
Making use of the relation of Eq.~\eqref{eq:green_relation} for the product of $G(k)G(k+q)$ we obtain
\begin{align}
\label{eq:cancel_1}
    \Pi_{SE_1} + \Pi_{SE_2} &= - \frac{1}{i q_0} \int \frac{d^3 k}{{(2 \pi)}^3} \left\{ \left(G^2(k)\Sigma(k)
    - G^2(k+q) \Sigma(k+q)\right) + G(k)G(k+q) \left(\Sigma(k+q) - \Sigma(k)\right)\right\} \bm{v}_{\bm{k}}^2\\
    &= - \frac{1}{i q_0} \int \frac{d^3 k}{{(2 \pi)}^3}  G(k)G(k+q) \left(\Sigma(k+q) - \Sigma(k)\right)
    \bm{v}_{\bm{k}}^2\,,
\end{align}
with the first term in parentheses in Eq.~\eqref{eq:cancel_1} is seen to cancel out when shifting variables $k+q \to
k$.

The MT diagram of Fig.~\ref{fig:MT} is given by
\begin{equation}
    \Pi_{MT} = \frac{\bar{g}^2}{\chi_0^2} \int \frac{d^3 k d^3 k' d^3 p}{{(2\pi)}^9} D^2(\bm{p})
    f^2_{\bm{k}+\frac{\bm{p}}{2}} f^2_{\bm{k}'+\frac{\bm{p}}{2}} G(k)G(k+q)G(k+p)G(k+p+q)G(k')G(k'+p)
    \, \bm{v}_{\bm{k}} \cdot \bm{v}_{\bm{k} + \bm{p}}\,.
\end{equation}
We will represent it as $ \Pi_{MT} \equiv \Pi_{MT}^{(a)} + \Pi_{MT}^{(b)}$, with
\begin{equation}
    \begin{split}
        \Pi_{MT}^{(a)} &= \frac{\bar{g}^2}{\chi_0^2} \int \frac{d^3 k d^3 k' d^3 p}{{(2\pi)}^9} D^2(\bm{p})
        f^2_{\bm{k}+\frac{\bm{p}}{2}} f^2_{\bm{k}'+\frac{\bm{p}}{2}}
        G(k)G(k+q)G(k+p)G(k+p+q)G(k')G(k'+p) \, \bm{v}_{\bm{k}}^2\\
        &= - \int \frac{d^3 k}{{(2\pi)}^3} G(k)G(k+q) \Gamma(k+q;q)\, \bm{v}_{\bm{k}}^2
        = - \int \frac{d^3 k}{{(2\pi)}^3}  G(k)G(k+q) \frac{\Sigma(k) - \Sigma(k+q)}{i q_0}\, \bm{v}_{\bm{k}}^2 \\
        &= - \left(\Pi_{SE_1} + \Pi_{SE_2}\right)\,,
    \end{split}
\end{equation}
which means that the sum of $\Pi_{MT}^{(a)}+\Pi_{SE_1} + \Pi_{SE_2}$ exactly cancels. The remaining part of
the MT diagram $\Pi_{MT}^{(b)}$ with the use of Eq.~\eqref{eq:green_relation} can be transformed as
\begin{equation}
	\label{Eq:app-ref-1}
    \begin{split}
        \Pi_{MT}^{(b)} ={}& \frac{\bar{g}^2}{\chi_0^2} \int \frac{d^3 k d^3 k' d^3 p}{{(2\pi)}^9} D^2(\bm{p})
        f^2_{\bm{k}+\frac{\bm{p}}{2}} f^2_{\bm{k}'+\frac{\bm{p}}{2}}
        G(k)G(k+q)G(k+p)G(k+p+q)G(k')G(k'+p)\, \bm{v}_{\bm{k}} \cdot (\bm{v}_{\bm{k} + \bm{p}} -\bm{v}_{\bm{k}})\\
        ={}& \frac{1}{{(i q_0)}^2}\frac{\bar{g}^2}{\chi_0^2} \int \frac{d^3 k d^3 k' d^3 p}{{(2\pi)}^9} D^2(\bm{p})
        f^2_{\bm{k}+\frac{\bm{p}}{2}} f^2_{\bm{k}'-\frac{\bm{p}}{2}} \left\{2G(k)G(k+q) - G(k)G(k+p+q)
        - G(k+q)G(k+p)\right\}\\
		&{} \times G(k')G(k'-p) \, \bm{v}_{\bm{k}} \cdot (\bm{v}_{\bm{k} + \bm{p}} -\bm{v}_{\bm{k}})\,,
    \end{split}
\end{equation}
with the change in integration variable from \(k'\) to \(k' -p\) in the last equality.

The $AL_1$ diagram coming from Fig.~\ref{fig:AL1} equals
\begin{equation}
	\label{Eq:app-ref-2}
    \begin{split}
        \Pi_{AL_1} ={}& \frac{\bar{g}^2}{\chi_0^2} \int \frac{d^3 k d^3 k' d^3 p}{{(2\pi)}^9} D^2(\bm{p})
        f^2_{\bm{k}+\frac{\bm{p}}{2}} f^2_{\bm{k}'-\frac{\bm{p}}{2}} G(k)G(k+q)G(k+p+q)G(k')G(k'+q)
        G(k'-p) \, \bm{v}_{\bm{k}} \cdot \bm{v}_{\bm{k}'}\\
        ={}& \frac{1}{{(i q_0)}^2}\frac{\bar{g}^2}{\chi_0^2} \int \frac{d^3 k d^3 k' d^3 p}{{(2\pi)}^9} D^2(\bm{p})
        f^2_{\bm{k}+\frac{\bm{p}}{2}} f^2_{\bm{k}'-\frac{\bm{p}}{2}} G(k+p+q) G(k'-p)\\
		&{}\times \left\{G(k)G(k') +G(k'+q)G(k+q) - G(k)G(k'+q) - G(k')G(k+q)\right\}
		 \, \bm{v}_{\bm{k}} \cdot \bm{v}_{\bm{k}'}\\
		 ={}& \frac{1}{{(i q_0)}^2}\frac{\bar{g}^2}{\chi_0^2} \int \frac{d^3 k d^3 k' d^3 p}{{(2\pi)}^9} D^2(\bm{p})
		f^2_{\bm{k}+\frac{\bm{p}}{2}} f^2_{\bm{k}'-\frac{\bm{p}}{2}} G(k') G(k'-p)\\
		&{}\times \left\{G(k)G(k+q+p) +G(k+p)G(k+q) - 2G(k)G(k+p)\right\}
		 \, \bm{v}_{\bm{k}} \cdot \bm{v}_{\bm{k}'}\,,
    \end{split}
\end{equation}
where in transitioning from the first to the second line, we used Eq.~\eqref{eq:green_relation}, and when
transitioning to the third equality, we performed a change of integration variables according to $k' \to k' -q$ and
$p \to p-q$ in both the second and third terms in braces, and $k \to k-q$ in the last term. These changes of
variables are justified because the free bosonic propagator $D(\bm{p})$ is frequency-independent, given that $q$ has
only non-zero frequency component.

Similarly, the $AL_2$ diagram coming from Fig.~\ref{fig:AL2} equals
\begin{equation}
	\label{Eq:app-ref-3}
    \begin{split}
        \Pi_{AL_2} ={}& \frac{1}{{(i q_0)}^2}\frac{\bar{g}^2}{\chi_0^2} \int \frac{d^3 k d^3 k' d^3
        p}{{(2\pi)}^9} D^2(\bm{p}) f^2_{\bm{k}+\frac{\bm{p}}{2}} f^2_{\bm{k}'-\frac{\bm{p}}{2}} G(k')
        G(k'-p)\\
		&{}\times \left\{2G(k)G(k+p) -G(k+p)G(k+q) -G(k)G(k+q+p) \right\}
		 \, \bm{v}_{\bm{k}} \cdot \bm{v}_{\bm{k}'-\bm{p}}\,.
    \end{split}
\end{equation}

Hence, the sum of the diagrams is given by (cf. Eq.~(A.10) in Ref.~\cite{Maslov_2017} without form-factors)
\begin{equation}
    \begin{split}
        \Pi_{MT}^{(b)}+\Pi_{AL_1}+\Pi_{AL_2} ={}& \frac{1}{{(i q_0)}^2}\frac{\bar{g}^2}{\chi_0^2} \int
        \frac{d^3 k d^3 k' d^3 p}{{(2\pi)}^9} D^2(\bm{p}) f^2_{\bm{k}+\frac{\bm{p}}{2}}
        f^2_{\bm{k}'-\frac{\bm{p}}{2}} G(k') G(k'-p)\\
		&{}\times \left\{2G(k)G(k+p) -G(k+p)G(k+q) -G(k)G(k+q+p) \right\}
		 \, \bm{v}_{\bm{k}} \cdot (\bm{v}_{\bm{k}+\bm{p}}+\bm{v}_{\bm{k}'-\bm{p}} -\bm{v}_{\bm{k}} -\bm{v}_{\bm{k}'})\,.
    \end{split}
\end{equation}
What remains to get the Eq.~\eqref{eq:polarization} is to perform some symmetrization of the indices. Let us make a
change of integration variables according to $k \to k -p/2$ and $k' \to k'+p/2$. Then
\begin{equation}
	\label{Eq:sum_diagram}
    \begin{split}
        \Pi_{MT}^{(b)}+\Pi_{AL_1}+\Pi_{AL_2} ={}& \frac{1}{{(i q_0)}^2}\frac{\bar{g}^2}{\chi_0^2} \int
        \frac{d^3 k d^3 k' d^3 p}{{(2\pi)}^9} D^2(\bm{p}) f^2_{\bm{k}} f^2_{\bm{k}'} G(k'+\tfrac{p}{2})
        G(k'-\tfrac{p}{2})\\
		&{}\times \left\{2G(k-\tfrac{p}{2})G(k+\tfrac{p}{2}) -G(k-\tfrac{p}{2}+q)G(k+\tfrac{p}{2})
		-G(k-\tfrac{p}{2})G(k+\tfrac{p}{2}+q) \right\}
		 \, \bm{v}_{\bm{k}-\tfrac{p}{2}} \cdot \Delta \bm{v}\,,
    \end{split}
\end{equation}
with $\Delta \bm{v}$ given by Eq.~\eqref{eq:velocity_change}.

Let us transform the first term in the braces by changing integration variables according to $k \leftrightarrow  k'$
and $p \to -p$.
The corresponding part of Eq.~\eqref{Eq:sum_diagram} becomes
\begin{equation}
	\label{Eq:sum_diagram_1}
    \begin{split}
       \Sigma \Pi^{I} ={}& \frac{1}{{(i q_0)}^2}\frac{\bar{g}^2}{\chi_0^2} \int \frac{d^3 k d^3 k' d^3 p}{{(2\pi)}^9}
       D^2(\bm{p}) f^2_{\bm{k}} f^2_{\bm{k}'} 2 G(k'+\tfrac{p}{2})
       G(k'-\tfrac{p}{2})G(k+\tfrac{p}{2})G(k-\tfrac{p}{2}) \, \bm{v}_{\bm{k}'+\tfrac{p}{2}} \cdot \Delta \bm{v}\,.
    \end{split}
\end{equation}
The remaining terms are transformed as
\begin{equation}
	\label{Eq:sum_diagram_2}
    \begin{split}
		\Sigma \Pi^{II} ={}& \frac{1}{q_0^2}\frac{\bar{g}^2}{\chi_0^2} \int \frac{d^3 k d^3 k' d^3 p}{{(2\pi)}^9}
		D^2(\bm{p}) f^2_{\bm{k}} f^2_{\bm{k}'} G(k'+\tfrac{p}{2})
		G(k'-\tfrac{p}{2})\left\{G(k-\tfrac{p}{2}+q)G(k+\tfrac{p}{2}) +G(k+\tfrac{p}{2}+q)G(k-\tfrac{p}{2}) \right\}
		\, \bm{v}_{\bm{k}-\tfrac{p}{2}} \cdot \Delta \bm{v}\\
		={}& \frac{1}{q_0^2}\frac{\bar{g}^2}{\chi_0^2} \int \frac{d^3 k d^3 k' d^3 p}{{(2\pi)}^9} D^2(\bm{p})
		f^2_{\bm{k}} f^2_{\bm{k}'}\, \bm{v}_{\bm{k}-\tfrac{p}{2}} \cdot \Delta \bm{v}\\
		&{}\times \left\{G(k'+\tfrac{p}{2}+\tfrac{q}{2}) G(k'-\tfrac{p}{2}+\tfrac{q}{2})
		G(k-\tfrac{p}{2}+\tfrac{q}{2})G(k+\tfrac{p}{2}-\tfrac{q}{2}) +G(k'+\tfrac{p}{2}+\tfrac{q}{2})
		G(k'-\tfrac{p}{2}+\tfrac{q}{2}) G(k+\tfrac{p}{2}+\tfrac{q}{2})G(k-\tfrac{p}{2}-\tfrac{q}{2}) \right\}\\
		={}& \frac{1}{q_0^2}\frac{\bar{g}^2}{\chi_0^2} \int \frac{d^3 k d^3 k' d^3 p}{{(2\pi)}^9} D^2(\bm{p})
		f^2_{\bm{k}} f^2_{\bm{k}'} G(k+\tfrac{p}{2})
		G(k-\tfrac{p}{2})\left\{G(k'+\tfrac{p}{2}+q)G(k'-\tfrac{p}{2}) +G(k'-\tfrac{p}{2}+q)G(k'+\tfrac{p}{2}) \right\}
		\, \bm{v}_{\bm{k}-\tfrac{p}{2}} \cdot \Delta \bm{v}\\
		={}& \frac{1}{q_0^2}\frac{\bar{g}^2}{\chi_0^2} \int \frac{d^3 k d^3 k' d^3 p}{{(2\pi)}^9} D^2(\bm{p})
		f^2_{\bm{k}} f^2_{\bm{k}'} G(k'+\tfrac{p}{2})
		G(k'-\tfrac{p}{2})\left\{G(k-\tfrac{p}{2}+q)G(k+\tfrac{p}{2}) +G(k+\tfrac{p}{2}+q)G(k-\tfrac{p}{2}) \right\}
		\, \bm{v}_{\bm{k}'+\tfrac{p}{2}} \cdot \Delta \bm{v}\,.
    \end{split}
\end{equation}
Moving to the second line, we adjusted the integration variables as \(k \to k-q/2\) and \(k' \to k'+q/2\). The shift
to the third line comes from applying \(p \to p+q\) in the first term and \(p \to p-q\) in the final term within the
braces. The concluding equivalence arises by swapping \(k \leftrightarrow k'\) and setting \(p \to -p\).

Summing up the Eqs.~(\ref{Eq:sum_diagram}--\ref{Eq:sum_diagram_2}), we finally obtain
\begin{equation}
\label{leading}
    \begin{split}
		\Pi_{MT}^{(b)}+\Pi_{AL_1}+\Pi_{AL_2} ={}& \frac{1}{2 q_0^2}\frac{\bar{g}^2}{\chi_0^2} \int \frac{d^3
		k d^3 k' d^3 p}{{(2\pi)}^9} D^2(\bm{p}) f^2_{\bm{k}} f^2_{\bm{k}'} G(k'+\tfrac{p}{2})
		G(k'-\tfrac{p}{2})\\
		&{}\times \left\{G(k-\tfrac{p}{2}+q)G(k+\tfrac{p}{2}) +G(k+\tfrac{p}{2}+q)G(k-\tfrac{p}{2})
		-2G(k-\tfrac{p}{2})G(k+\tfrac{p}{2}) \right\}
		\, (\bm{v}_{\bm{k}-\tfrac{p}{2}}+\bm{v}_{\bm{k}'+\tfrac{p}{2}}) \cdot \Delta \bm{v}\\
		={}& \frac{1}{4 q_0^2}\frac{\bar{g}^2}{\chi_0^2} \int \frac{d^3 k d^3 k' d^3 p}{{(2\pi)}^9} D^2(\bm{p})
		f^2_{\bm{k}} f^2_{\bm{k}'} G(k'+\tfrac{p}{2}) G(k'-\tfrac{p}{2})\\
		&{}\times \left\{2G(k-\tfrac{p}{2})G(k+\tfrac{p}{2}) -G(k-\tfrac{p}{2}+q)G(k+\tfrac{p}{2})
		-G(k+\tfrac{p}{2}+q)G(k-\tfrac{p}{2})  \right\}
		\,{(\Delta \bm{v})}^2\,,
    \end{split}
\end{equation}
where the transition to the last line involves the change of $p \to -p$, taking into account that $\Delta \bm{v}$ is
an odd function of $p$. The integration over the frequencies $\int dk_0 dk'_0 dp_0 \ldots$ is elementary, given the
simple pole structure of the free electron's Green function Eq.~\eqref{eq:el_propagator}, leading to
Eq.~\eqref{eq:polarization}.
\end{widetext}

\bibliography{paper}

\end{document}